# The Solar Dynamics Observatory in the Living With a Star Era: From Solar Observation to Predictive Heliophysics

**Madhulika Guhathakurta**

*Heliophysics Division, Science Mission Directorate, NASA Headquarters, Washington, DC, USA*

*Email: madhulika.guhathakurta@nasa.gov*

*ORCID: 0000-0001-5357-4452*

---

**Abstract**

The Solar Dynamics Observatory (SDO), launched in 2010 as part of NASA's Living With a Star (LWS) program, represents a methodological transition in heliophysics: from identifying discrete solar events to characterizing the continuously evolving state of the solar atmosphere as the upstream boundary of a coupled Sun-heliosphere-geospace system. By providing co-registered, high-cadence, full-disk measurements of the solar magnetic field via the Helioseismic and Magnetic Imager (HMI), coronal emission across multiple temperature regimes via the Atmospheric Imaging Assembly (AIA), and solar extreme-ultraviolet irradiance via the Extreme Ultraviolet Variability Experiment (EVE), SDO has enabled the solar atmosphere to be treated as a dynamical system evolving in time rather than as a catalog of isolated events. This capability has direct consequences for space-weather specification, ionospheric and thermospheric modeling, heliospheric transport, and the radiation environment relevant to human exploration beyond low Earth orbit. The mission's open-data policy has produced a uniform, continuously calibrated archive that has contributed to more than 8 400 peer-reviewed publications, enabling statistical, model-driven, and machine-learning approaches that depend on long, uninterrupted observational records. The Space-weather HMI Active Region Patches (SHARPs), derived from near-real-time HMI vector magnetograms, are used directly in operational flare probability forecasting and serve as the standard training dataset for machine-learning prediction systems; AIA near-real-time imagery is used routinely for solar activity monitoring and CME identification in forecasting environments worldwide. This perspective traces SDO's scientific lineage, infrastructure role, and the emerging applications of its archive to data-driven approaches and foundation models in the service of predictive heliophysics.

**Keywords:** Solar Dynamics Observatory -- heliophysics -- space weather -- solar magnetic field -- coronal imaging -- extreme-ultraviolet irradiance -- data-driven methods -- predictive heliophysics

---

## 1. Introduction

The term heliophysics was introduced in the mid-2000s to define an interdisciplinary discipline encompassing the Sun, the heliosphere, and their interaction with planetary environments. The intent was to move beyond the traditional boundaries of solar, magnetospheric, and ionospheric physics and to treat the Sun-heliosphere-geospace domain as a single coupled physical system. Within this framework, solar variability must be understood not only as a subject of solar physics but as the driver of the space environment that affects technological systems and human activity in space. The first volume of the Heliophysics textbook series was published by Cambridge University Press in 2009 (Schrijver and Siscoe, 2009).

The Living With a Star (LWS) program preceded this terminology and was formulated to address the coupled-system problem directly: to investigate the physical processes linking solar variability to the heliosphere and geospace, and to establish the observational and modeling foundations needed for space-environment specification. In doing so, LWS provided both the programmatic and scientific rationale for the system-science approach that heliophysics would later name, formalize, and broaden as a discipline.

Earlier missions, particularly SOHO (Domingo, Fleck, and Poland, 1995), established that coronal mass ejections (CMEs), irradiance variability, and solar-wind structures originate at the Sun and propagate through the heliosphere to Earth. These discoveries demonstrated the causal connection between solar activity and space-weather effects and made quantitative Sun-Earth connection studies possible. SOHO carried a broad suite of spectral diagnostic instruments, but it was not designed to deliver the specific combination that SDO provides: full-disk vector magnetic-field measurements, simultaneous multi-temperature coronal imaging at high cadence and spatial resolution, and co-registered irradiance spectroscopy, all from a single platform without time-sharing among instruments.

Addressing this limitation required observations capable of tracking magnetic and thermal structure everywhere on the Sun with sufficient cadence and stability to describe the solar atmosphere as a continuously evolving system. Such measurements are necessary if solar observations are to be used as upstream input for heliospheric transport calculations, geospace simulations, and the specification of the space environment.

The Solar Dynamics Observatory (SDO), launched in 2010 as part of the LWS program, was designed to meet this requirement (Pesnell, Thompson, and Chamberlin, 2012). By providing co-registered, high-cadence measurements of the solar atmosphere and magnetic field over more than a solar cycle, SDO made it possible to treat solar variability as the evolving state of a dynamical system rather than as a sequence of isolated events. This capability allows solar observations to be used to constrain models of the heliosphere and geospace and forms an essential observational foundation for predictive heliophysics.

In this sense, SDO represents not only a major observational mission but a methodological transition in the field: from identifying disturbances after they occur to characterizing the time-dependent environment in which they develop.

## 2. Personal Perspective: Building a Mission of Continuity

The Solar Dynamics Observatory did not arise from the need for higher spatial resolution alone. It originated from a scientific limitation: solar activity could be described observationally, but not followed as the continuous evolution of a global magnetic system.

The LWS program introduced an explicit requirement to understand the Sun as part of a coupled Sun-heliosphere-geospace system whose variability affects technological society. Within this framework, solar variability could no longer be treated as a sequence of isolated events, but had to be measured as the evolving state of the solar atmosphere. This requirement exposed an observational gap that earlier missions had not been designed to address.

Missions of the SOHO era transformed the field by providing sustained observations of the solar atmosphere and heliosphere. Measurements of coronal structure, irradiance variability, and solar-wind outflow established CMEs, large-scale magnetic fields, and radiative variability as the drivers of space-weather disturbances at Earth. These observations made it possible to associate solar activity with geospace response and marked the beginning of quantitative Sun-Earth connection studies.

Yet an important limitation remained. Although observations were continuous, they did not provide uniform, high-cadence, full-disk coverage across multiple diagnostics sufficient to follow the evolution of magnetic systems everywhere on the Sun simultaneously. Solar activity could often be observed once it became prominent, but the gradual development of magnetic complexity and the buildup of free energy leading to instability were only partially sampled.

As the objectives of the LWS program expanded from discovery to specification of the space environment, the problem had to be reformulated. If solar variability defines the conditions in which technological systems operate, the relevant quantity is not the occurrence of individual events but the evolving global state of the solar atmosphere. Meeting this objective required observations capable of following magnetic and thermal structure over the entire solar disk without interruption (Pesnell, Thompson, and Chamberlin, 2012).

SDO was designed to provide this temporal completeness. High-cadence, co-registered measurements of the vector magnetic field (HMI), coronal emission across multiple temperature regimes (AIA), and solar irradiance (EVE) allow the solar atmosphere to be followed continuously over the entire disk. Processes occurring on minute timescales, including flux emergence, magnetic reconnection, and localized heating, can be tracked as stages of development rather than inferred retrospectively.

As continuous observations became available, the corona appeared persistently active, structured by ongoing magnetic reorganization and small-scale energy release across the entire disk. Questions shifted from identifying eruptions to understanding how instability develops. The broader implication became evident as the data began to be used beyond solar physics: continuous solar monitoring provided upstream information for heliospheric and geospace modeling, and the distinction between solar observation and environmental monitoring began to blur.

The change introduced by SDO is therefore not continuity alone, but temporal completeness. The buildup phases of eruptions, the accumulation of free energy in active regions, and the slow reorganization of large-scale field patterns all unfolded at cadences and over spatial scales that no prior mission had been designed to follow continuously across the full disk. SDO allows the evolving state of the solar atmosphere to be tracked everywhere on the disk at cadence sufficient to resolve the buildup of instability. This capability is essential for treating the Sun as the time-dependent boundary condition of the heliosphere and geospace.

## 3. The Lineage: From Discovery to Evolution -- SOHO, STEREO, and SDO

Progress in heliophysics has proceeded through successive advances in observational capability, each allowing the Sun-heliosphere-geospace system to be understood at a deeper physical level. Rather than replacing earlier missions, later observatories extended the range of questions that could be addressed, moving from discovery of solar drivers, to reconstruction of heliospheric propagation, to continuous measurement of the evolving solar state.

The SOHO mission established the Sun as the origin of heliospheric disturbances through continuous observation of the corona, solar wind, and solar irradiance (Domingo, Fleck, and Poland, 1995). Measurements of CMEs, large-scale magnetic structure, and radiative variability demonstrated that solar activity drives changes throughout the heliosphere and produces the space-weather effects observed at Earth. These observations made it possible to associate disturbances in geospace with specific solar sources and marked the beginning of quantitative Sun-Earth connection studies. SOHO carried a wide range of spectral and particle instruments, but it was not built to deliver the particular combination SDO provides: vector rather than line-of-sight magnetic-field maps, simultaneous full-disk imaging across multiple temperature diagnostics at 12-second cadence, and co-registered irradiance measurements, without time-sharing among instruments.

The STEREO mission addressed a different and equally fundamental problem: the inability to see the Sun and the interplanetary medium from multiple vantage points simultaneously (Kaiser et al., 2008). Operating as a pair of spacecraft drifting ahead of and behind Earth in its orbit, STEREO provided the first continuous observations of the full 360 degrees of solar longitude, and together with SOHO and SDO formed a fleet capable of seeing nearly every part of the solar surface at once (Guhathakurta, 2013). The three-dimensional structure and propagation of CMEs could be reconstructed directly rather than inferred from projection effects, and solar storms could be tracked through the inner heliosphere regardless of their direction of travel. This capability fundamentally changed space weather forecasting from an Earth-centric discipline to an interplanetary one: when active region AR1429 rotated across the far side of the Sun in March 2012, firing eruptions in every direction, mission operators throughout the solar system knew the storms were coming. What STEREO did not provide was the sustained, co-registered monitoring of the solar atmosphere needed to characterize the buildup of instability before eruptions occur.

The Solar Dynamics Observatory was designed to provide this capability. By combining high-cadence full-disk imaging, vector magnetic-field measurements, and irradiance observations, SDO allows the solar atmosphere to be observed as a time-dependent system. Instead of identifying eruptions only after they become visible, the gradual evolution of magnetic topology, the accumulation of free energy, and the onset of instability can be followed directly.

In this progression, the roles of the three missions are complementary: SOHO established the solar origin of heliospheric disturbances; STEREO extended the view to the full 360 degrees of solar longitude and made space weather forecasting interplanetary; SDO made it possible to follow the formation and buildup of those disturbances in time. Together, these capabilities provide the observational foundation required to treat solar variability as the time-dependent boundary condition of the heliosphere and geospace, completing the transition from event-based solar physics to system-level heliophysics.

This observational lineage continues with Parker Solar Probe (Fox et al., 2016), Solar Orbiter (Muller et al., 2020), and PUNCH (DeForest et al., 2026), for which SDO provides essential coronal context. In-situ measurements of the solar wind and energetic particles acquired close to the Sun require knowledge of the coronal conditions from which those structures originated, a requirement that SDO's continuous full-disk coverage is uniquely suited to address. PUNCH, designed to image the continuous flow of solar wind from the outer corona into the inner heliosphere, extends this context into the domain where coronal structures transition to interplanetary ones. Cross-instrument studies combining SDO with Parker Solar Probe, Solar Orbiter, and PUNCH represent an increasingly central direction in heliospheric research.

## 4. The Observatory as Infrastructure

SDO was designed as a continuous observatory rather than a point-and-observe mission. Achieving uniform, high-cadence, full-disk measurements across multiple diagnostics required not only stable instruments but an operational architecture capable of uninterrupted observation, persistent downlink, and consistent calibration over long intervals. The ground segment therefore functions as part of the instrument, ensuring that the data record remains temporally complete and internally consistent (Pesnell, Thompson, and Chamberlin, 2012).

The three SDO instruments provide complementary measurements that together characterize the solar atmosphere as an evolving physical system. AIA provides full-disk images at 0.6 arcsecond pixel scale and 4 096 by 4 096 resolution across seven extreme-ultraviolet and two ultraviolet passbands, with a nominal cadence of 12 seconds per channel (Lemen et al., 2012). This combination of spatial resolution, temporal cadence, and multi-temperature coverage allows the corona to be observed simultaneously from chromospheric to flare temperatures, making it possible to follow the development of coronal structures in both space and time. HMI provides full-disk vector magnetograms, Dopplergrams, and continuum intensity images with 0.5 arcsecond pixel scale, at 45-second cadence for line-of-sight measurements and 12-minute cadence for vector fields (Schou et al., 2012; Scherrer et al., 2012). EVE measures solar spectral irradiance from 0.1 to 105 nm with spectral resolution sufficient to resolve individual emission lines (Woods et al., 2012). EVE incorporates an autonomous flare campaign mode that increases observing cadence in response to detected solar activity without requiring ground intervention, allowing it to capture the detailed spectral evolution of solar flares in real time, with direct relevance to operational space-weather monitoring.

The co-registration of these diagnostics allows the relationships between magnetic evolution, energy release, and radiative output to be followed continuously over the entire disk. Because these measurements are obtained at uniform cadence and with stable calibration, they can be combined to describe the evolving state of the solar atmosphere rather than isolated features.

The operational concept of SDO was driven by the requirements of system-level heliophysics. Many processes relevant to the Sun-heliosphere-geospace connection develop over timescales ranging from minutes to the solar cycle and involve interactions between widely separated regions of the Sun. Observations that are intermittent, non-uniform, or limited to selected targets cannot capture this evolution. Continuous full-disk coverage, together with a telemetry and ground-processing system capable of sustaining high data rates over extended periods, was therefore essential to the mission design.

The resulting data system produces a uniform, continuously calibrated record of the solar atmosphere extending over more than a solar cycle. Because the observations are acquired with stable cadence and consistent processing, the archive can be used not only for individual investigations but also for long-duration statistical studies, model validation, and data-driven analysis. In this sense, the SDO archive functions as scientific infrastructure: a persistent observational reference that supports both physical modeling and empirical approaches to the Sun-heliosphere-geospace system (Figure 1).

## 5. Scientific Phase Transition: What Became Observable with SDO

The primary scientific impact of the Solar Dynamics Observatory was not the discovery of entirely new classes of solar phenomena, but the ability to follow the evolution of the solar atmosphere with sufficient continuity, cadence, and diagnostic coverage to study solar activity as a time-dependent physical system. Earlier missions revealed the existence of eruptions, waves, and irradiance variability, but the processes leading to these disturbances were often inferred from incomplete temporal coverage or from observations obtained in a limited number of diagnostics. The uniform, co-registered measurements provided by SDO made it possible to investigate how magnetic and thermal structure evolves across the entire disk prior to the onset of instability.

One example is the formation of eruptive active regions. Prior to SDO, the gradual evolution of magnetic topology, flux emergence, and shearing motions leading to instability could not be followed uniformly over the full solar disk. With the continuous vector magnetic-field measurements provided by HMI, the buildup of magnetic shear, helicity, and free energy can be tracked throughout the lifetime of an active region (Schou et al., 2012; Sun et al., 2012; DeRosa et al., 2015). These measurements allow the magnetic field to be treated as a time-dependent boundary condition for models of coronal and heliospheric dynamics, rather than as a sequence of isolated snapshots. Helioseismic diagnostics further extend this capability by revealing subsurface flows and far-side activity, linking the evolution of surface magnetic structure to processes occurring beneath the photosphere (Schou et al., 2012; Lindsey and Braun, 2000).

High-cadence imaging from AIA has similarly changed the interpretation of coronal structure. Observations across multiple extreme-ultraviolet passbands show that the corona is not a steady atmosphere punctuated by occasional events, but a continuously evolving system structured by small-scale, intermittent energy release occurring over a wide range of spatial and temporal scales (Lemen et al., 2012; Testa et al., 2014; Klimchuk, 2015). Full-disk coverage at 12-second cadence makes it possible to observe this evolution as a globally connected process: disturbances in one region propagate along field lines to distant parts of the solar surface, and the large-scale magnetic topology shapes where and how energy is released (Schrijver and Title, 2011). The ability to follow loops, brightenings, and waves across the entire disk simultaneously, rather than in isolated patches, has been essential for testing models of coronal heating and magnetic reconnection that depend on both local energy release and global field connectivity.

Solar irradiance variability provides another example of the importance of simultaneous, multi-diagnostic observations. Measurements from EVE show that radiative output changes on timescales ranging from minutes to the solar cycle, with significant consequences for the ionosphere and thermosphere (Woods et al., 2012; Chamberlin, Woods, and Eparvier, 2012; Qian et al., 2010). EVE's continuous, spectrally resolved coverage revealed classes of solar variability that had been invisible to previous broadband instruments. Among the most significant was the discovery of gradual, long-duration EUV enhancements associated with C-class flares that, despite their modest GOES classification, deliver substantial energy to the thermosphere over extended periods (Woods et al., 2011). These events would have gone undetected by GOES X-ray monitoring alone, and their thermospheric impact would have been unattributed. Interpreting such variations requires knowledge of the magnetic and coronal structures from which the emission originates. The co-registered measurements provided by HMI, AIA, and EVE allow changes in irradiance to be related directly to the evolution of active regions and coronal emission, enabling quantitative studies of Sun-ionosphere coupling that were not possible with earlier datasets.

The importance of temporal completeness is also evident in studies of filament formation, flux transport, and polar-field evolution. These processes develop over months to years and involve interactions between widely separated regions of the Sun. Continuous, uniform observations over more than a solar cycle allow the global magnetic field to be followed without interruption, revealing departures from simple periodic behavior and providing constraints on models of the solar dynamo and solar-cycle variability (Hathaway, 2015; Jiang et al., 2014; Petrie, 2013). Direct measurements of meridional circulation from HMI Dopplergrams revealed a more complex multi-layered flow structure than previously known, with implications for flux transport and dynamo models (Zhao et al., 2013). The long, uninterrupted record of full-disk Dopplergrams has also enabled the detection of large-scale inertial oscillations, including equatorial Rossby waves, that were previously inaccessible to observation and whose properties constrain models of the solar interior (Loptien et al., 2018).

Across these examples, the change introduced by SDO is consistent. Solar activity can be observed not only when it becomes obvious, but throughout the evolution that leads to it. Magnetic structure, coronal emission, and radiative output can be followed continuously over the full disk, allowing solar variability to be described as a trajectory in time rather than as a sequence of isolated events. This capability is essential for the goals of system-level heliophysics: because solar radiation, energetic particles, and magnetic structure define the conditions throughout the heliosphere and in geospace, prediction requires knowledge of the solar state before disturbances occur.

## 6. From the Sun to the Heliosphere and Geospace: A Coupled Physical System

Solar variability reaches Earth's environment through three principal pathways: electromagnetic radiation, magnetized plasma, and energetic particles. Each pathway responds to different aspects of solar activity, and each depends on knowledge of the evolving solar state for quantitative specification. Radiation, magnetized plasma, and energetic particles originating at the Sun determine the conditions encountered by spacecraft, influence the Earth's upper atmosphere, and modulate the flux of galactic cosmic rays throughout the heliosphere. Because these disturbances propagate outward from the Sun at different speeds and along different trajectories, understanding and predicting their effects requires continuous knowledge of the solar state, not just at the moment of eruption but throughout the buildup that precedes it.

Solar extreme-ultraviolet and X-ray emission ionizes and heats the upper atmosphere, modifying electron density in the ionosphere and expanding the thermosphere. Both effects propagate into practical consequences: degraded GPS accuracy, disrupted high-frequency radio communication, and increased aerodynamic drag on low Earth orbit satellites. Specification of these responses requires continuous irradiance measurements linked to the magnetic and coronal structures that produce them (Woods et al., 2012; Chamberlin, Woods, and Eparvier, 2012; Qian et al., 2011), a combination that EVE, HMI, and AIA provide jointly. The availability of uniform, spectrally resolved irradiance measurements together with co-registered magnetic and coronal observations has enabled improved specification of ionospheric response and supported efforts to apply data-driven and machine-learning techniques to Sun-ionosphere coupling.

Magnetized plasma carried by the solar wind provides another channel through which solar variability influences geospace. The speed and structure of the solar wind, the orientation of the interplanetary magnetic field, and the properties of CMEs are determined by the evolving magnetic configuration of the solar atmosphere. These quantities define the upstream boundary conditions for heliospheric transport models and for simulations of magnetosphere-ionosphere coupling. In operational practice, synoptic magnetograms from HMI drive potential-field source-surface extrapolations that define the coronal boundary conditions for the WSA-Enlil model (Arge and Pizzo, 2000; Odstrcil, 2003), the primary physics-based solar wind forecast tool used at NOAA's Space Weather Prediction Center. More broadly, modern modeling frameworks rely on HMI synoptic maps, flux-transport models, and data assimilation techniques to incorporate solar observations into global simulations of the heliosphere and geospace environment (Linker et al., 1999; Riley et al., 2015; Fisher et al., 2015).

The Space-weather HMI Active Region Patch (SHARP) data series gives this operational chain a specific and widely used form. SHARPs are automatically extracted time series of vector magnetic-field data and derived parameters, including total unsigned flux, magnetic shear angle, current helicity, and non-potential energy proxies, centered on individual active regions as they cross the solar disk (Bobra and Couvidat, 2015; Hoeksema et al., 2014). Operational forecasters at NOAA's Space Weather Prediction Center use SHARP parameters directly for flare probability estimation, and SHARPs are the standard training data for machine-learning flare prediction systems. They are perhaps the clearest existing example of an SDO data product that bridges basic research and operational practice.

The Gannon Superstorm of 2024 May illustrated the value of continuous magnetic monitoring at a scale not seen in years. NOAA Active Regions 13664 and 13668 produced a sequence of major eruptions that drove the most intense geomagnetic disturbance of Solar Cycle 25. HMI observations captured the magnetic-field evolution over the preceding days: rapid flux emergence among the highest rates recorded during the SDO mission, delta-type sunspot development, and the accumulation of substantial free magnetic energy in the days before the first major flare (Sun et al., 2024). The event showed directly that the precursory evolution detectable in photospheric field data has both scientific and operational significance.

Solar energetic particles (SEPs) accelerated in flares and at CME-driven shocks present a third pathway, one with particular relevance for astronaut safety and for high-latitude aviation. SEP onset can precede the development of geomagnetic disturbances, compressing the available warning time. The particle fluence and energy spectrum depend on the magnetic geometry of the eruption site and the large-scale heliospheric field, making continuous solar observation a prerequisite for physically based SEP forecasting (Reames, 2013; Desai and Giacalone, 2016).

Galactic cosmic-ray flux at Earth is modulated over the solar cycle by the evolving heliospheric field, and transiently suppressed by Forbush decreases during major geomagnetic events (Cane, 2000; Potgieter, 2013). Interpreting these variations requires knowledge of the global solar magnetic configuration and its long-term evolution, available from the HMI synoptic archive.

The Solar Dynamics Observatory provides a key component of the observational capability required to specify these pathways. Continuous full-disk measurements from HMI, AIA, and EVE supply magnetic, coronal, and radiative inputs used in synoptic mapping, global MHD models, and data-assimilation frameworks that describe the heliosphere and geospace environment. These data have become part of the infrastructure supporting both research and operational space-weather modeling, enabling solar variability to be incorporated directly into simulations of the coupled Sun-heliosphere-geospace system.

## 7. Exploration Beyond Low Earth Orbit: Solar Variability as an Environmental Constraint

Human activity in low Earth orbit occurs largely within the protection of Earth's magnetosphere, where the effects of solar variability are partially shielded by the geomagnetic field and atmosphere. Missions beyond low Earth orbit lose that protection. For crewed missions to the Moon or Mars, the radiation environment is determined primarily by SEP events, galactic cosmic rays, and the state of the heliospheric field, all of which vary with solar activity and cannot be anticipated without continuous solar monitoring.

SEP events are the acute hazard. Particle fluxes can rise by orders of magnitude within minutes to hours of a large eruption, and the severity of an event depends on the magnetic topology of the eruption site and the connectivity between the source region and the spacecraft location (Reames, 2013; Desai and Giacalone, 2016). Probabilistic forecasting of SEP occurrence, and deterministic forecasting of onset time and spectrum once an eruption is detected, both depend on the kind of real-time magnetic and coronal data that SDO provides.

Solar radiation affects spacecraft and planetary upper atmospheres over a wider range of timescales. Flare-driven extreme-ultraviolet and X-ray enhancements heat and expand planetary ionospheres within minutes. Cycle-scale irradiance changes alter atmospheric density at orbital altitudes and affect aerobraking calculations. EVE and AIA observations have extended knowledge of both short-term variability and cycle-scale trends (Woods et al., 2012; Qian et al., 2011).

Galactic cosmic rays, modulated by the large-scale heliospheric field, represent the chronic radiation background for interplanetary missions. Their flux increases during solar minimum and decreases during active periods, varying by factors of two or more over the cycle (Cane, 2000; Potgieter, 2013). Accurate assessment of integrated crew dose on long-duration missions requires knowledge of the solar-cycle phase and the evolving heliospheric field, both of which are tracked using HMI synoptic magnetograms.

Disturbances that affect radiation exposure, plasma conditions, and communication environments originate in the solar atmosphere and propagate through the heliosphere on timescales comparable to operational decision cycles. Specification of the exploration environment therefore requires continuous observations of the Sun together with models that incorporate solar measurements as boundary conditions for heliospheric and radiation-transport calculations. Planning for crewed exploration depends on solar monitoring at a level similar to meteorological monitoring for terrestrial operations. SDO's continuous observations of the magnetic field, corona, and irradiance constitute part of the sensor infrastructure that exploration operations will require.

## 8. Open Data as Scientific Infrastructure

SDO data have been publicly available from the start of the mission, with no proprietary period. The rationale was practical as well as principled: the science questions that motivated SDO required global, long-duration observations analyzed by a wide community, and restricting access to small teams would have limited the return on the investment in the mission. The same logic that drove the instrument design drove the data policy.

Making a uniform, continuously calibrated, full-disk dataset freely available changed how solar physics is practiced. Investigators worldwide analyze the same observations, so discrepancies between studies can be attributed to differences in analysis methodology rather than differences in input data. Competing physical interpretations can be evaluated against a common observational record (Pesnell, Thompson, and Chamberlin, 2012; Lemen et al., 2012; Schou et al., 2012). By mid-2025 the SDO data archive had supported more than 8 400 peer-reviewed publications with over 150 000 citations, a measure less of the mission's size than of its role as shared infrastructure for the field (Figure 1c).

The multi-cycle record has made possible studies that require long baselines. Solar-cycle evolution, active-region flux statistics, irradiance variability on cycle timescales, and polar-field trends can now be examined over a sample large enough to yield statistically meaningful results (Hathaway, 2015; DeRosa et al., 2012; Woods et al., 2012). These are problems that could not be approached with data from any earlier mission.

Outside solar physics, HMI synoptic magnetograms have become standard inputs to coronal and solar-wind models. Potential-field source-surface extrapolations, MHD coronal models, and heliospheric transport codes all ingest photospheric field maps as boundary conditions (Arge and Pizzo, 2000; Linker et al., 2017; Riley et al., 2015; Hayashi, 2015). Common solar inputs allow model outputs from different groups to be compared and combined in ways that were not previously feasible.

The same open archive has enabled machine-learning applications to solar physics. Training and validating data-driven models requires long, consistently calibrated, well-documented datasets accessible without restriction. The SDO record meets these requirements, and has supported a growing body of work on flare prediction, irradiance forecasting, feature detection, and representation learning (Bobra and Couvidat, 2015; Camporeale, 2019; Nishizuka et al., 2018). The scale of the archive, combined with its consistency across years and its simultaneous coverage of multiple physical quantities, is what makes these applications tractable.

In this sense, the SDO archive functions not only as a collection of observations but as scientific infrastructure: a persistent observational reference that supports collaborative analysis, cross-domain modeling, and data-driven approaches required to understand a system whose behavior cannot be captured by isolated measurements alone. The transition from event-based studies to system-level heliophysics depends not only on instrumentation but on the existence of a shared, persistent observational record. The SDO archive represents the first resource of this kind for the Sun.

## 9. Data-Driven Heliophysics: From Archive to Representation

The SDO archive differs from earlier solar datasets not just in volume but in character. Observations at fixed cadence, consistent calibration, and co-registered diagnostics across more than a solar cycle preserve the temporal relationships between magnetic field, coronal structure, and irradiance that earlier datasets could not. A record with these properties can support analyses that go beyond identifying events or fitting models to individual cases: it can be used to infer how the solar atmosphere evolves, and to construct representations of that evolution applicable to prediction.

Data-driven methods take advantage of this character directly. Rather than requiring a first-principles physical model to close the system, they learn the relationships embedded in the data. Flare probability estimates derived from SHARP parameters, for example, do not solve the equations of magnetohydrodynamics; they extract statistical associations between measurable field properties and subsequent activity from a large sample of observed active regions (Bobra and Couvidat, 2015; Camporeale, 2019). The validity of the approach depends on having a training sample large enough and consistent enough to be representative, which is precisely what the SDO archive provides.

Foundation models extend this principle. The Surya Heliophysics Foundation Model (Roy et al., 2025; Munoz-Jaramillo et al., 2025), trained on the SDO archive, learns general representations of solar variability that can be adapted to tasks including flare prediction, irradiance reconstruction, and solar-wind boundary specification. The architecture is similar to large language models in natural language processing: a single model trained on a large, diverse dataset provides a base representation that is then fine-tuned for specific applications. Surya demonstrates that the SDO archive is large and uniform enough to support this approach.

Physics-based and data-driven methods are complementary. Physical models define the equations governing plasma dynamics and constrain the allowable evolution of the system. Data-driven models learn from the observed record what trajectories the system actually takes within those constraints. Hybrid methods that combine physical equations with data assimilation and statistical learning, applied to irradiance specification, thermospheric density modeling, and heliospheric boundary conditions, draw on both (Fisher et al., 2015; Woods et al., 2012). The quality of these hybrid approaches depends on the temporal completeness and diagnostic coverage of the observational record. SDO provides both.

The broader implication is that the SDO archive is not merely a repository of solar images. It is a record of how the solar atmosphere evolves, structured in a way that allows that evolution to be analyzed, modeled, and ultimately predicted. The transition from descriptive solar physics to predictive heliophysics depends on having such a record. SDO is the first solar mission to produce one at the required scale.

## 10. Living With a Star: Toward Predictive Heliophysics

SDO occupies a specific position in the development of heliophysics. It did not establish the field's foundational results, which came from earlier missions. It did not resolve the detailed physics of any single phenomenon. What it did was make the solar atmosphere continuously observable, everywhere on the disk, at the cadence and in the diagnostics needed to treat solar variability as a dynamical system. That change in observational capability enabled a change in what kinds of questions can be asked and addressed.

The LWS program framed the objective that SDO was built to serve: understanding the chain of processes from solar activity to its effects on technological systems and human activity in space. Meeting that objective requires solar observations that go beyond detecting events. It requires knowledge of the evolving solar state before disturbances form and propagate into the heliosphere. The gap between identifying eruptions after they occur and anticipating them from precursory observations is exactly the gap that SDO's continuous, multi-diagnostic monitoring was designed to close.

Over more than a solar cycle, the SDO data record has been incorporated into heliospheric transport models, geospace simulations, and operational space-weather forecasting. Its open-data policy has made it the primary training resource for machine-learning approaches to solar activity prediction. Its cadence and diagnostic coverage have enabled the development of data-driven representations of solar variability broad enough to be applied across multiple prediction tasks.

The emergence of foundation models trained on the SDO archive points toward a longer-term aspiration. Roy et al. (2025) and Munoz-Jaramillo et al. (2025) demonstrate that a single large-scale representation of solar variability, learned from the full multi-cycle record, can be adapted across prediction tasks that previously required separate models and separate datasets. As the archive grows and modeling approaches mature, such representations may eventually support the kind of continuous, system-level specification of the heliospheric environment that the LWS program originally envisioned: not the prediction of individual events, but the ongoing characterization of the variable star we live with.

The work is not finished. SDO provides the solar boundary condition, but connecting that boundary condition to conditions at Earth, at other planetary bodies, and throughout the inner heliosphere requires advances in transport modeling, data assimilation, and the treatment of uncertainty. The mission's contribution to this larger program is to have made the solar atmosphere measurable as a continuous physical system. Whether that measurement can be turned into reliable prediction of conditions throughout the heliosphere is the central challenge of the field that SDO helped define.

---

## Acknowledgments

This paper is dedicated to Alan Title and Karel Schrijver, whose scientific vision and decades of contributions shaped the field this work describes. Both were close colleagues and giants of solar physics whose loss is deeply felt.

The Solar Dynamics Observatory is supported by NASA's Living With a Star program. The Joint Science Operations Center (JSOC) at Stanford University and the Laboratory for Atmospheric and Space Physics (LASP) at the University of Colorado provide essential data processing and archive infrastructure. The author acknowledges the SDO instrument teams and the broader heliophysics community whose work this article draws upon.

The author is a civil servant at NASA Headquarters and has served as SDO's Program Scientist from the mission's inception. No external funding was received for the preparation of this manuscript.

## Declarations

**Funding:** No external funding was received for the preparation of this manuscript. The author's position is supported by NASA Headquarters.

**Competing interests:** The author is a co-author on Roy et al. (2025), which is cited in this article. No other competing interests are declared.

**Author contributions:** M. Guhathakurta conceived, wrote, and revised the manuscript in its entirety.

**Data availability:** All SDO data referenced in this article are publicly available at the Joint Science Operations Center (https://jsoc.stanford.edu) and through the NASA Heliophysics Digital Resource Library (https://helio.data.nasa.gov/mission/SDO). No new data were generated for this perspective article.

---

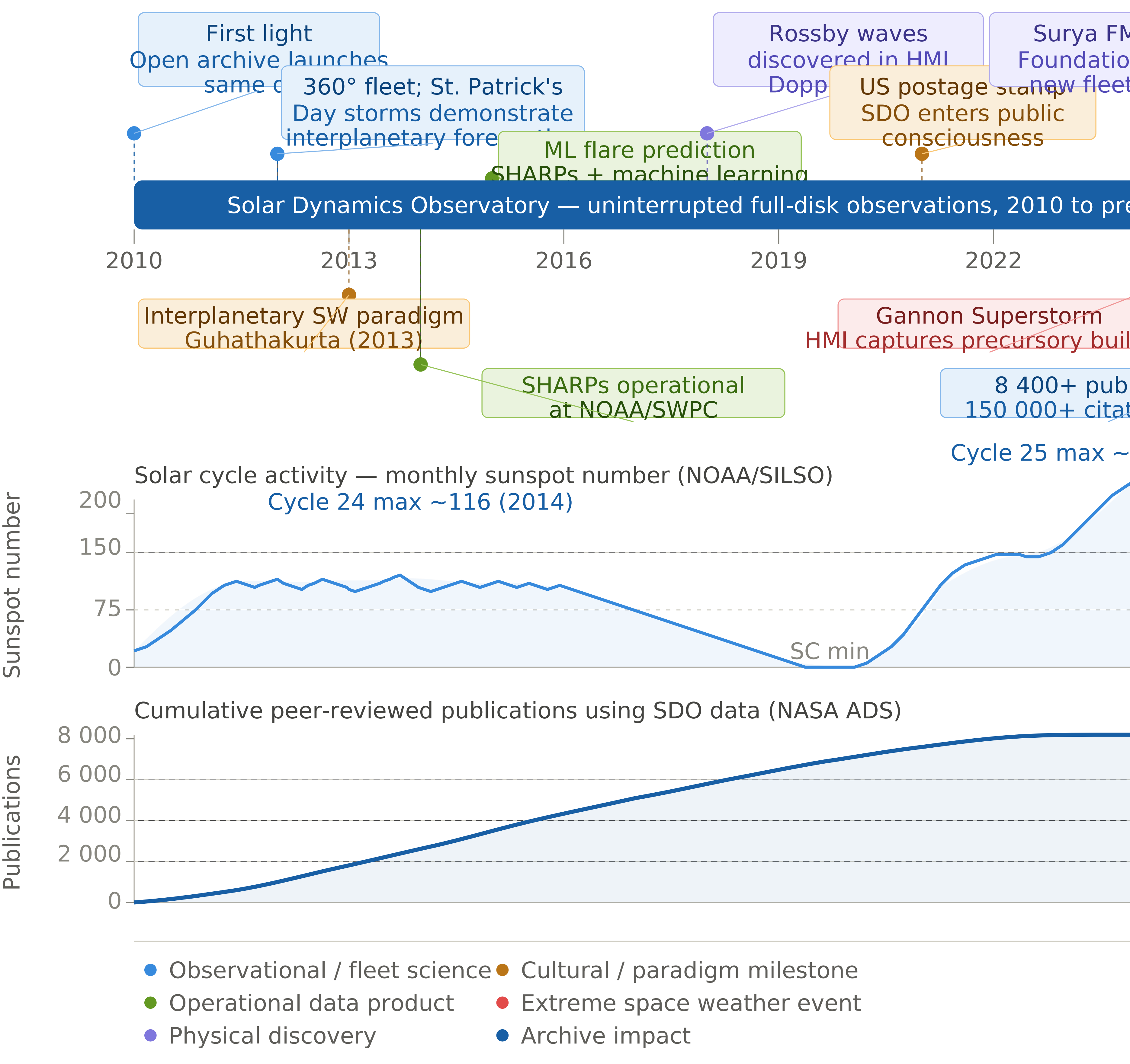
SDO: Observational Timeline, Milestones, and Scientific Impact
First light
Open archive launches
360° fleet; St. Patrick's
Day storms demonstrate
Rossby waves
discovered in HMI
US postage stamp
SDO enters public
consciousness
ML flare prediction
Solar Dynamics Observatory — uninterrupted full-disk observations, 2010 to pre
2010
2013
2016
2019
2022
Interplanetary SW paradigm
Guhathakurta (2013)
Gannon Superstorm
SHARPs operational
at NOAA/SWPC
Cycle 25 max ~
Solar cycle activity — monthly sunspot number (NOAA/SILSO)
Cycle 24 max ~116 (2014)
Sunspot number
200
150
75
0
SC min
Cumulative peer-reviewed publications using SDO data (NASA ADS)
Publications
8 000
6 000
4 000
2 000
0
Observational / fleet science
Cultural / paradigm milestone
Operational data product
Extreme space weather event
Physical discovery
Archive impact
Sunspot data: NOAA/SWPC and SILSO, Royal Observatory of Belgium. Publication count: NASA Ast

## Figure Legends

**Figure 1.** The Solar Dynamics Observatory as scientific infrastructure. (a) Timeline of uninterrupted SDO operations from 2010 to 2025, annotated with ten milestones spanning five dimensions of mission impact: observational and fleet science (blue); operational data products (green); physical discoveries enabled by the long data record (purple); cultural and paradigm milestones (amber); and extreme space weather events demonstrating operational value (red). (b) Monthly sunspot number for Solar Cycles 24 and 25 (NOAA/SWPC; SILSO, Royal Observatory of Belgium), illustrating that SDO has operated across a full solar cycle and into the maximum of the next. The vertical dashed line marks the Gannon Superstorm of 2024 May, the most intense geomagnetic disturbance of Solar Cycle 25. (c) Cumulative count of peer-reviewed publications using SDO data, reaching more than 8 400 by mid-2025 (NASA Astrophysics Data System). The accelerating and then sustained growth reflects the mission's role as shared scientific infrastructure rather than a single-investigator resource.